\documentstyle[aps,prl,epsf,multicol]{revtex}

\draft
\begin{document}

\title{Robust half-metallic ferromagnetism in the zincblende CrSb}
\author{Bang-Gui Liu}
\address{
Institute of Physics \& Center of Condensed Matter Physics,
Chinese Academy of Sciences, Beijing 100080, P. R. China}
\date{\today}
\maketitle

\widetext

\begin{abstract}
Using the accurate first-principle method within
density-functional theory, we systematically study CrSb in the
zincblende (zb) structure. The zb CrSb is predicted of robust
half-metallic ferromagnetism (HMFM) with a magnetic moment of
$3.000 \mu_B$ per formula. It is much better than other zb
compounds with HMFM because its spin-flip gap reaches 0.774 eV at
the equilibrium volume and persists nonzero with its volume
changing theoretically from $-21\%$ to $+60\%$. It is found there
may be a common mechanism for the HMFM in all the zb Cr- and
Mn-pnictides. Since being compatible with the $I\!I\!I$-$V$
semiconductors, this excellent HMFM of the zb CrSb should be
useful in spin electronics and other applications.
\end{abstract}


\raggedcolumns
\begin{multicols}{2}
\narrowtext

{\it Introduction.} Half-metallic (HM) ferromagnets are absorbing
more and more attention around the world, because they have only
one electronic spin channel at the Fermi energy and should act as
ideal components of spintronics devices\cite{sptr,HMM}. Since de
Groot {\it et al} first predicted HM ferromagnetism in Heusler
compounds\cite{heusler,groot2} in 1983, several HM ferromagnets,
such as NiMnSb\cite{NiMnSb}, CrO$_2$\cite{CrO2},
Fe$_3$O$_4$\cite{Fe3O4}, and the manganate materials\cite{LSMO},
have been theoretically predicted and then experimentally
confirmed. Much effort has been paid to understand the mechanism
behind the HM magnetism and to study its implication in various
physical properties\cite{phys}. At the same time it is still very
important to find new HM ferromagnets which are more promising for
basic properties and applications.\cite{groot1}

Recently, much theoretical and experimental attention is paid to
the HM ferromagnetism in the zincblende (zb) compounds such as
$M$As and $M$Sb ($M$ is a transition-metal
element)\cite{epi-mnx,mnasdot,cras,cras1,crsb,crsb1}, which are
structurally and chemically compatible with the important
$I\!I\!I$-$V$ and $I\!I$-$V\!I$ semiconductors. The zb MnAs was
theoretically predicted to be of so-called `nearly HM'
ferromagnetism\cite{nhmm}. Its Fermi level crosses the
majority-spin (MAS) energy bands but touches the bottom of the
minority-spin (MIS) conduction bands, and therefore there is no
gap for spin-flip excitations in this case. Accurate calculations
within density-functional theory (DFT) showed that the zb MnSb and
MnBi are of true HM ferromagnetism\cite{xu}. On the other hand, it
was already found that the zb CrAs is a true HM ferromagnet with a
finite spin-flip gap \cite{cras,shirai}. Although crystalizing
into the NiAs (na) phases in bulk form \cite{MnX}, the zb phases
of CrAs\cite{cras,cras1}, CrSb\cite{crsb} and MnAs\cite{mnasdot}
have be successfully fabricated in form of thin films, multilayers
or nanostructures on the $I\!I\!I$-$V$ semiconductors. It is very
interesting to find new HM ferromagnets with not only the same
crystalline structure but also much more promising properties.

In this paper we systematically study CrSb in the zb structure
with the accurate full-potential (linear) augmented plane wave
plus local orbitals (FLAPWLO) method within density-functional
theory, and thereby predict that it is a true HM ferromagnet with
a magnetic moment of $3.000 \mu_B$ per formula. Being much more
advantageous than the zb CrAs and Mn-pnictides, its theoretical
spin-flip gap reaches 0.774 eV at equilibrium volume and persists
nonzero with its volume changing from $-21\%$ to $+60\%$. Because
not only the zb CrAs films of 5-6 unit cells (30 \AA) in thickness
\cite{cras,cras1} but also the zb CrSb thin films of 2 unit cells
in thickness\cite{crsb} have been successfully fabricated by means
of epitaxial growth, the high-quality zb ferromagnetic (FM) CrSb
films (and multilayers) of more than 5 unit cells in
thickness\cite{crsb} are believed to be obtained in the same way
in the near future and therefore realize the prediction. It is
also found there may be a common mechanism, which is similar to
that of the Heusler by de Groot {\it et al}, for the HM
ferromagnetism in all the zb Cr- and Mn-pnictides. Because being
compatible with the $I\!I\!I$-$V$ semiconductors, the robust zb HM
FM phase of CrSb should be useful in spin electronics and in other
applications.

{\it Computational details.} We make use of the Vienna package
WIEN2k\cite{wien2k} for FLAPWLO method within density functional
theory (DFT)\cite{DFT} for all our calculations. As for the EC
potentials, we mainly took the PBE96\cite{PBE96} version for the
generalized gradient approximation (GGA), but other versions of
GGA and local spin density approximation (LSDA)\cite{PW92} were
also used for comparison and confirmation. The relativistic effect
was taken into account in the scalar style, but the spin-orbit
coupling was neglected in the results presented in this paper
because it has little effect on the main results. We always used
3000 k-points in the Brillouin Zone, took $R_{mt}*K_{max}$ as 8.0
and made the expansion up to $l=10$ in the muffin tins. The radii
$R_{mt}$ of the muffin tins was chosen to be approximately
proportional to the corresponding ionic radii and as large as
possible until reaching 2.3 (Cr) and 2.8 (Sb) Bohr, respectively.
They were kept equivalent for different phases at their
equilibrium volumes in order to accurately compare their total
energies. The self-consistent calculations were considered to be
converged only when the integrated charge distance per cell,
$\int{}|\rho_n-\rho_{n-1}|dr$, between input charge density
($\rho_{n-1}(r)$) and output ($\rho_{n}(r)$) was less than
0.00001.
\begin{figure}[tbp]
\epsfxsize=0.40\textwidth \epsfbox{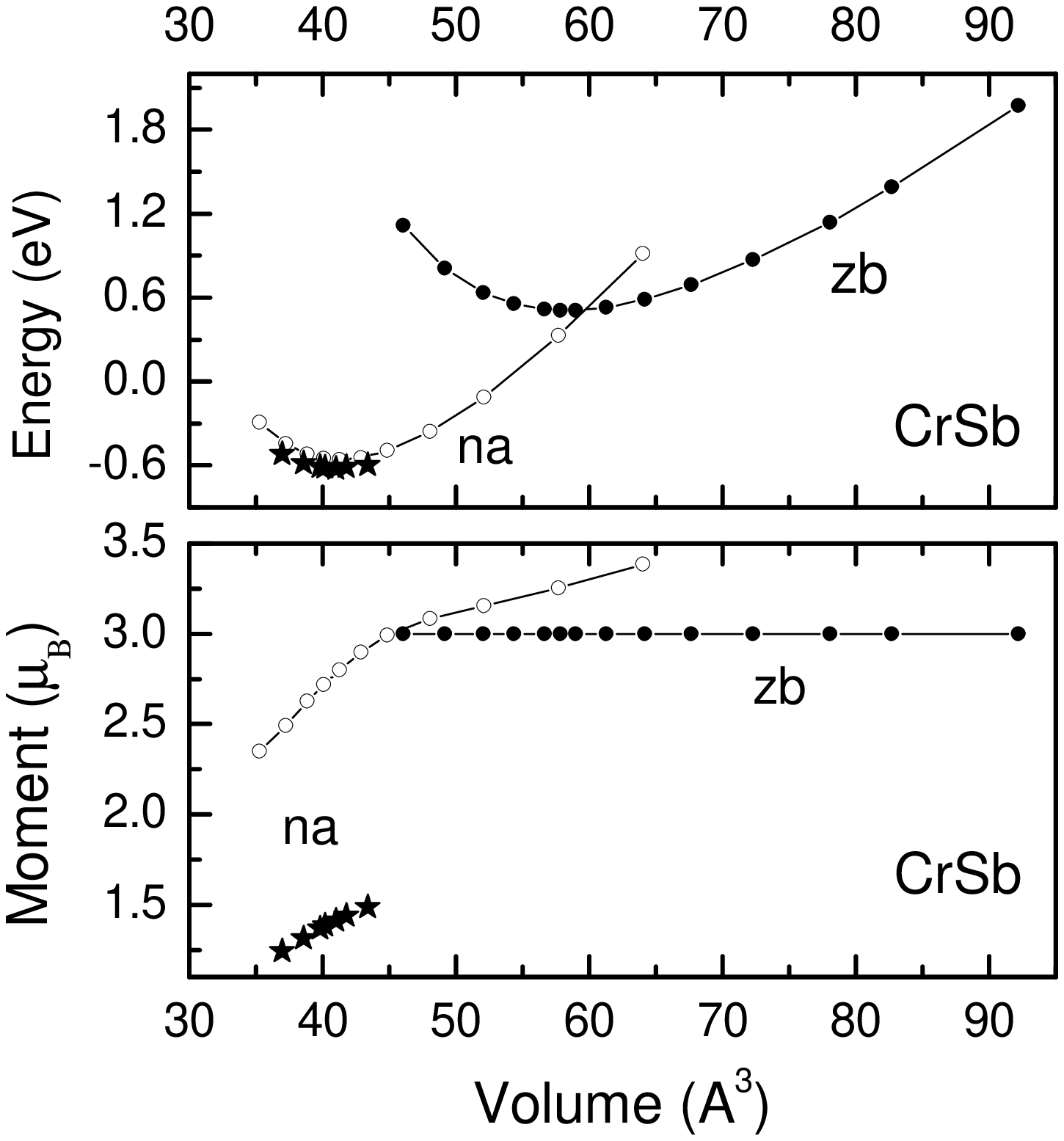} \caption{The
structure optimization (upper panel) and the magnetic moments per
formula (lower panel) as functions of the volume per formula. The
solid circles are for the zb FM phase, the open circles the na FM
phase, and the stars the na AFM phase. The (volume, moment) at the
equilibrium volume are equivalent to (57.825 \AA$^3$, 3.000
$\mu_B$), (41.275 \AA$^3$, 2.801 $\mu_B$), and (41.00 \AA$^3$,
1.412 $\mu_B$) for the three phases, respectively. The moment of
the zb phase remains the same 3.000 $\mu_B$ between -21 \% (46.0
\AA$^3$) and +60 \% (92.3 \AA$^3$) of the volume changes.}
\label{fig01}
\end{figure}

{\it Main findings.} At low temperature CrSb crystalizes into the
na antiferromagnetic (AFM) phase with the modulation vector along
the [001] direction. The AFM N\'eel temperature is 710 K and the
experimental lattice constants are $a=4.122$ \AA{} and $c=5.464$
\AA. Although not being the ground state, the zb phase of CrSb has
been grown epitaxially on the substrates of GaAs, (Al,Ga)Sb and
GaSb, and proved to be of FM order\cite{crsb} with Curie
temperature higher than 400 K. We made systematical structure
optimization of CrSb in the two phases: na and zb. The total
energy results as functions of formula volume are shown in the
upper panel of Fig. 1. The equilibrium na AFM phase is indeed
lowest in total energy and therefore is the ground state of CrSb,
with their lattice constants being the same as the experimental
result\cite{crsb1}. The na FM phase is also studied for
comparison. Its equilibrium energy is a little higher than the AFM
phase, being consistent with the AFM exchange coupling. The
equilibrium zb FM phase is about 1 eV higher in total energy than
the na AFM phase, and therefore should not exist as bulk crystals.
Anyway, it has been grown as thin films epitaxially on the
$I\!I\!I$-$V$ semiconductors\cite{crsb}. In the lower panel of
Fig. 1 the dependence of the magnetic moments on the formula
volume are presented. The moment increases with the volume for
both the na phases, but the moment of the zb phase remains the
same 3.000 $\mu_B$ at least between -21 \% and 60 \% in the
relative volume changes. This characteristic of the zb phase
implies that the zb CrSb is an HM ferromagnet.
\begin{figure}[btp]
\epsfxsize=0.45\textwidth \epsfbox{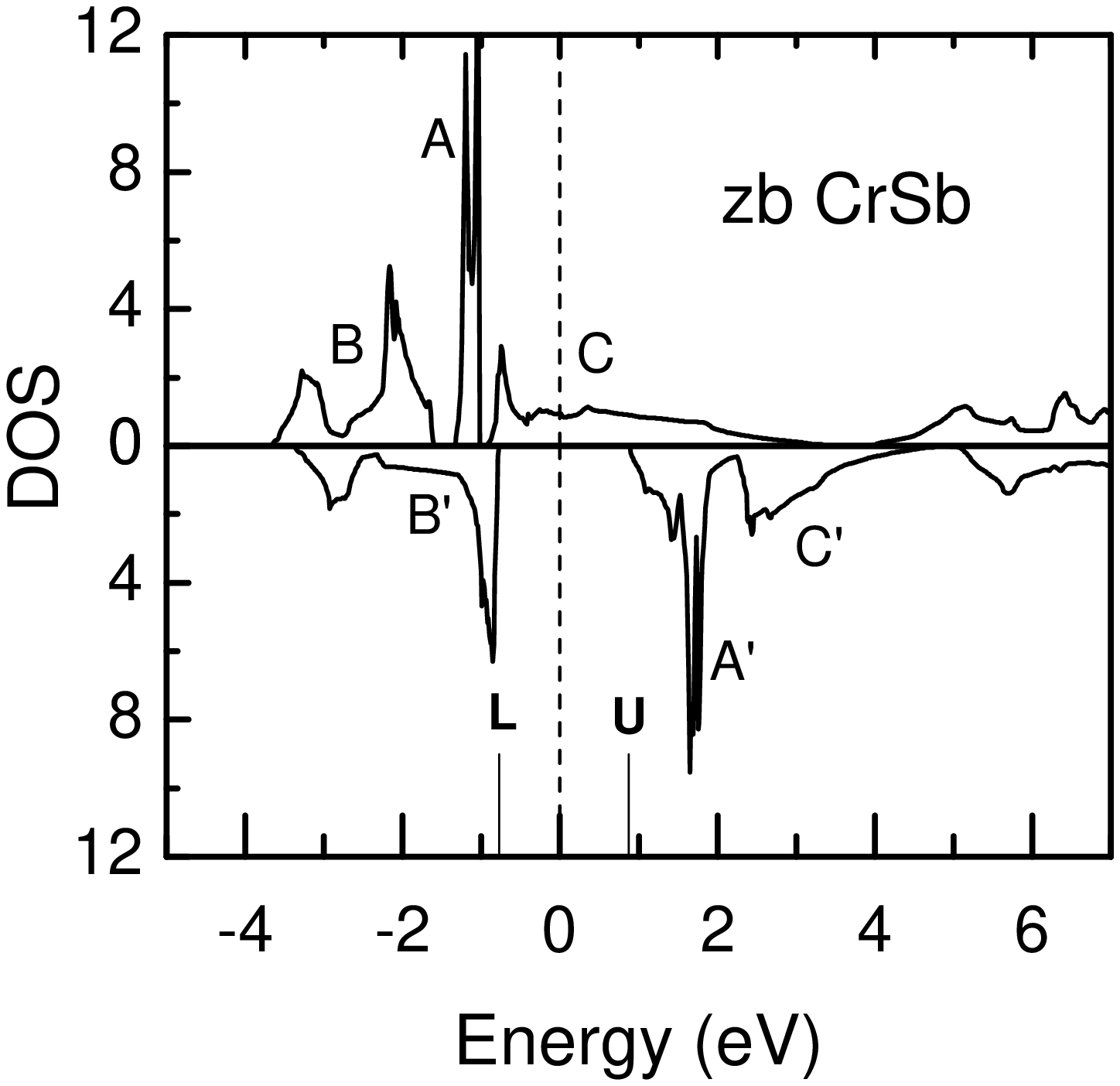} \caption{The
spin-dependent total density of states (DOS) of the equilibrium zb
phase. The upper part denotes the majority-spin (MAS) DOS and the
lower part the minority-spin (MIS) DOS. The bottom (U) of the MIS
conduction bands is 0.872 eV and the top (L) of the MIS valence
bands -0.774 eV, with the spin-flip gap being 0.774 eV.}
\label{fig02}
\end{figure}

In Fig. 2 we present the spin-dependent total DOS for the zb CrSb
phase at its equilibrium volume. The upper part is for the MAS
total DOS between -5 and 7 eV and the lower part the MIS. It is
clear that the MAS electrons are metallic but there is a gap of
1.646 eV around the Fermi energy for the MIS electronic bands. The
bottom (U) of the MIS conduction bands is at +0.872 eV and the top
(L) of the valence MIS bands at -0.774 eV. The gap for creating an
MIS hole at the top of the MIS valence bands by exciting an MIS
electron into the conducting MAS bands is 0.774 eV and the gap for
an MIS electron at the bottom of the MIS conduction bands is 0.872
eV. As a result, the minimal energy gap for a spin-flip
excitation, or the HM gap, is 0.774 eV. This nonzero gap is
essential to form a true HM ferromagnet. Fig. 3 demonstrates the
corresponding energy bands in the high-symmetrical directions.
There are 3 MIS and 5 MAS full-filled bands above -5 eV. The
$\Gamma_1$ bands, corresponding to the Cr s states, are pushed
above the Fermi energy by the interaction with the Sb s electrons.
The three filled MIS bands ($\Gamma_{15}$) and the three filled
MAS bands ($\Gamma_{15}$) reflect the bonding of the three Sb p
electrons and three of the Cr electrons, and do not contribute to
the total moment. The two very narrow MAS bands ($\Gamma_{12}$)
around -1 eV are mainly of Cr d $e_{g}$  character. The two MAS
bands crossing the Fermi energy originates from the Sb p and Cr d
$t_{2g}$ states, but their filled sections are mainly of Cr d
$t_{2g}$. The magnetic moment of 3.000 $\mu_B$ per formula results
from the remaining 3 Cr d electrons with 3 Cr electrons already
bonding with the 3 Sb p electrons. Therefore, it is shown by the
DOS, magnetic moment, and energy bands that the zb CrSb is a true
HM ferromagnet with the moment 3.000 $\mu_B$ per formula.
\begin{figure}[tbp]
\epsfxsize=0.45\textwidth \epsfbox{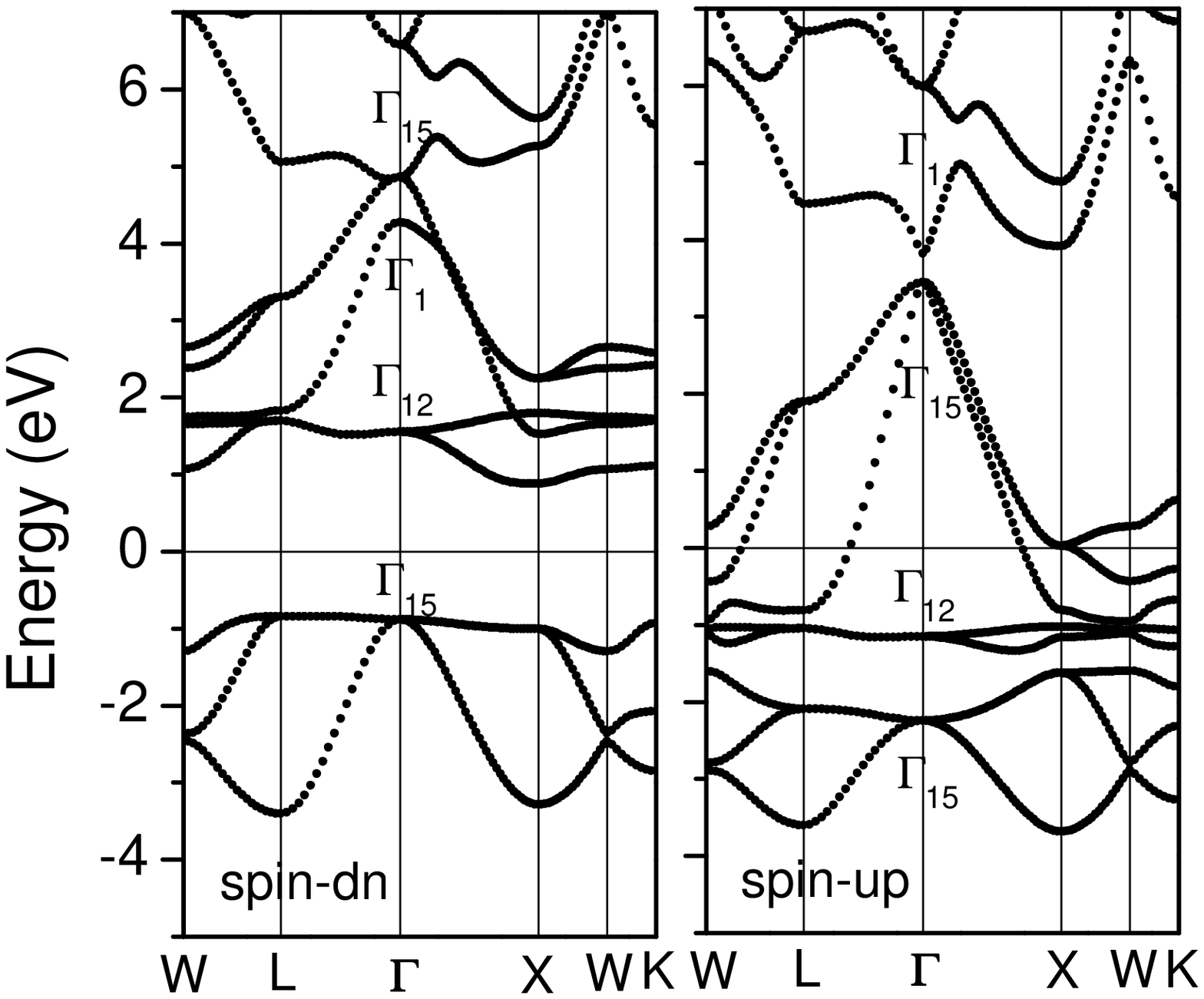} \caption{The
spin-dependent energy bands of the equilibrium zb phase. The right
panel is for the majority-spin (MAS) bands and the left the
minority-spin (MIS) bands. There are three full-filled MIS bands
and five full-filled MAS bands above -5 eV. There are only two
partly-filled MAS bands, which together contribute 1 $\mu_B$ to
the total moment per formula.} \label{fig03}
\end{figure}

{\it How does the HM ferromagnetism form?} The Cr s states have
moved above the Fermi energy because they must be orthogonal with
the Sb 5s states\cite{xu}. It is shown by comparing the
spin-dependent partial DOS, partial charge density and band
structures that the two triplet band clusters in B and B' both
result from the bonding of the Sb p and Cr d $t_{2g}$ triplet
electrons. There are approximately 2 and 2.6 Sb p (and 1 and 0.4
Cr d $t_{2g}$) electrons in B and B', respectively. On the other
hand, the Cr d $e_g$ states do not bond with Sb so that there
exist two very narrow MAS bands (A) at -1.15 eV and the two MIS
bands (A') around 1.7 eV. They are dominantly of Cr d $e_g$
character, and are separated 2.85 eV from each other due to the
exchange interaction. This exchange splitting is consistent with
$Im$ in the Stone theory if we take $I\approx 1$ eV for the
exchange integration and $m=3$ for the present case. This kind of
exchange splitting usually is enough to produce ferromagnetism in
traditional ferromagnets such as bcc Fe, but, in addition, other
energy bands must be of right structures at right positions in
order to form HM ferromagnetism. In the case of the zb CrSb, the
excellent HM ferromagnetism is formed because (1) The large
exchange splitting of the narrow $e_g$ bands pushes the MIS bands
in A' above the Fermi energy, opening the MIS gap at the Fermi
energy, and at the same time pulls the MAS $e_g$  bands in A below
the Fermi energy; (2) The presence of the Sb 5s electrons pushes
the extended Cr 4s states above the $e_g$ bands, keeping the MIS
gap clean through avoiding them from mixing with the Cr d bands;
(3) The triplet bands in C are partly filled with 1 electron per
formula in order to conserve the total number of the electrons.
The Cr $t_{2g}$ states form polarly covalent bonds with the Sb 5p
states so that approximate 1.5 electrons per formula are
transferred towards the Sb sites. This enhances the Cr d exchange
splitting and pins the Fermi energy at the right position  through
guaranteeing that the $e_g$ bands remain very narrow and the DOS
at the bottom of the bands in C is large enough. This explanation
is similar to that by de Groot {\it et al} in their
cases\cite{heusler,groot2}. The difference is mainly in the role
played by the Sb 5s electrons.

Comparing with the zb CrAs and the Mn pnictides, we find their
energy band structures to be similar to that of the zb CrSb. It is
observed in all these cases that (1) the $\Gamma_{1}$ bands are
above the narrow $\Gamma_{12}$ bands in band clusters A and A',
(2) the Fermi energies are in the MIS band gaps formed between the
empty MIS $\Gamma_{12}$ in A' and the full-filled MIS
$\Gamma_{15}$ bands in B', and (3) the corresponding MAS
$\Gamma_{15}$ in B and $\Gamma_{12}$ bands in A are all
full-filled. The difference relies only on the relative positions
of the $\Gamma_{1}$ bands and the $\Gamma_{15}$ bands in C and C'.
This means that the mechanism behind the HM ferromagnetism should
be the same for all the zb Cr- and Mn-pnictides. It appears to be
also the same as that in the Heusler compounds by de Groot {\it et
al}\cite{heusler,groot2}. It is very interesting that the
mechanism behind the half-metallic ferromagnetism is the same in
many cases, if not in all cases.

{\it Is the HM ferromagnetism theoretically robust?} The spin-flip
gap, the essential characteristic for HM ferromagnetism, persists
nonzero even when we theoretically change the cell volume from -21
\% and 60 \%. During this volume change, the MIS gap changes
little, but the spin-flip gap decreases with the cell deviating
from the equilibrium volume. The Fermi energy moves upwards when
the volume decreases, but downwards when the volume increases. Its
theoretical spin-flip gap, 0.774 eV, is much wider than 0.46 eV of
the zb CrAs, 0.2 eV of the zb MnSb, and 0 eV of the zb MnAs. Its
HM ferromagnetism survives from the large theoretical volume
changes, being much more robust than the zb CrAs between (-15 \%,
+27 \%) and the zb MnSb between (-6 \%, +12 \%). It is clearly
much better than other known HM or nearly HM ferromagnetism in the
zb compounds. We already confirmed the main results by using other
EC potentials and parameters. Because of physical errors in the
GGA potential, there should be some calculation errors, but the
relative volume error must be less than 10 \% in these accurate
full-potential DFT calculations. Anyway, since persisting for the
volume changes from -21\% to +60\%, the HM ferromagnetism in the
zb CrSb is indeed robust against any theoretical errors, even more
robust when taking into account the fact that all real DFT
calculations underestimate energy gaps.

{\it What does this accurate DFT prediction imply for real zb CrSb
materials?} Because the zb CrSb is theoretically 1 eV unstable
with respect to the ground-state phase, it has been realized only
in form of thin films through epitaxial growth on the
$I\!I\!I$-$V$ semiconductors\cite{crsb}. Anyway, the zb CrAs is
also theoretically 0.9 eV unstable with respect to the na CrAs
phase, but the zb CrAs thin films with 5-6 unit cells in thickness
\cite{cras} have been successfully fabricated. This success makes
us believe that one can successfully fabricate in the near future
the high-quality zb CrSb thin films with more than 5 unit cells in
thickness and similar zb multilayers. When the thickness of a zb
CrSb film sample is larger than 5 unit cells, the prediction,
based on the accurate DFT crystal calculation, must apply to their
internal parts deep beyond 2 unit cells, where there is very
little surface effect. The real spin-flip gap should be
approximately 1 eV because all LSDA and GGA calculations within
DFT usually underestimate the gap by 30 - 60 \%. A wide spin-flip
gap implies that nearly 100 \% spin polarization can be achieved
at quite high temperature. The zb CrSb of the HM ferromagnetism
may be the only candidate, which not only has excellent robust HM
ferromagnetism but also shares the same crystalline structure as
the important $I\!I\!I$-$V$ semiconductors, and therefore, should
be useful in spin electronics and in fabricating HM magnetic
quantum wells, dots and multilayers with some of them.

{\it In summary}, we systematically study CrSb in the zb structure
with the full-potential LAPW method within density-functional
theory, and thereby predict that it is a robust HM ferromagnet
with a magnetic moment of $3.000 \mu_B$ per formula. The large
exchange splitting of the narrow Cr $e_g$ bands is essential in
forming its ferromagnetism. The presence of the low-lying Sb 5s
electrons and the polar-covalent bonding between the Sb p and Cr
$t_{2g}$ states both are necessary in further forming its
half-metallic characteristic. It is found there may be a common
mechanism for the HM ferromagnetism in all the cases of the zb Cr-
and Mn-pnictides. Its excellent spin-flip gap makes it possible to
achieve nearly-full spin polarization at high temperature. The HM
ferromagnetic phase of the zb CrSb is much better than those of
the other zb compounds, and therefore should be useful in spin
electronics and other applications because being compatible with
the important $I\!I\!I$-$V$ semiconductors.

{\it Acknowledgment.} This work is supported in parts by China
National Key Projects of Basic Research (Grant No. G1999064509),
by Nature Science Foundation of China, and by British Royal
Society under a collaborating project with Chinese Academy of
Sciences.

\end{multicols}

\end{document}